\documentclass[sigconf]{acmart}

\usepackage{amsmath,amsfonts}
\usepackage{algorithmic}
\usepackage{graphicx}
\usepackage{textcomp}
\usepackage{xcolor}
\usepackage{fancyhdr}
\usepackage{hyperref}
\usepackage{subcaption}
\usepackage[normalem]{ulem}

\AtBeginDocument{%
	}

\def\fixme#1{\bgroup \color{red}{[{#1}]}\egroup}

\newcommand{\specialcellright}[2][r]{
	\begin{tabular}[#1]{@{}c@{}}#2\end{tabular}
}

\begin{document}
	
	\acmYear{2024}\copyrightyear{2024}
	\setcopyright{acmlicensed}
	\acmConference[HPDC '24]{International Symposium on High-Performance Parallel and Distributed Computing}{June 3--7, 2024}{Pisa, Italy}
	\acmBooktitle{International Symposium on High-Performance Parallel and Distributed Computing (HPDC '24), June 3--7, 2024, Pisa, Italy}
	\acmDOI{10.1145/3625549.3658686}
	\acmISBN{979-8-4007-0413-0/24/06}
	
	\title{Reinforcement Learning-based Adaptive Mitigation of Uncorrected DRAM Errors in the Field}

	
	\author{Isaac Boixaderas}
	\affiliation{%
		\institution{Barcelona Supercomputing Center}
		\streetaddress{P.O. Box 1212}
		\city{Barcelona}
		\country{Spain}
	}
	\email{isaac.boixaderas@bsc.es}
	
	\author{Sergi Mor\'e}
	\affiliation{%
		\institution{Barcelona Supercomputing Center}
		\city{Barcelona}
		\country{Spain}}
	\email{sergi.more@bsc.es}
	
	\author{Javier Bartolome}
	\affiliation{%
		\institution{Barcelona Supercomputing Center}
		\city{Barcelona}
		\country{Spain}}
	\email{javier.bartolome@bsc.es}
	
	\author{David Vicente}
	\affiliation{%
		\institution{Barcelona Supercomputing Center}
		\city{Barcelona}
		\country{Spain}}
	\email{david.vicente@bsc.es}
	
	\author{Petar Radojkovi\'c}
	\affiliation{%
		\institution{Barcelona Supercomputing Center}
		\streetaddress{1 Th{\o}rv{\"a}ld Circle}
		\city{Barcelona}
		\country{Spain}}
	\email{petar.radojkovic@bsc.es}
	
	\author{Paul M. Carpenter}
	\affiliation{%
		\institution{Barcelona Supercomputing Center}
		\city{Barcelona}
		\country{Spain}}
	\email{paul.carpenter@bsc.es}
	
	\author{Eduard Ayguad\'e}
	\affiliation{%
		\institution{Barcelona Supercomputing Center\\Universitat Politècnica de Catalunya}
		\city{Barcelona}
		\country{Spain}}
	\email{eduard.ayguade@bsc.es}
	
	\renewcommand{\shortauthors}{Boixaderas et al.}
	
	\begin{abstract}	
		Scaling to larger systems, with current levels of reliability, requires
		cost-effective methods to mitigate hardware failures. One of the main
		causes of hardware failure is an uncorrected error in memory, which
		terminates the current job and wastes all computation since the last
		checkpoint. This paper presents the first adaptive method for
		triggering uncorrected error mitigation. It uses a prediction approach
		that considers the likelihood of an uncorrected error and its current
		potential cost. The method is based on reinforcement learning, and the
		only user-defined parameters are the mitigation cost and whether the
		job can be restarted from a mitigation point. We evaluate our
		method using classical machine learning metrics together with a
		cost–benefit analysis, which compares the cost of mitigation actions
		with the benefits from mitigating some of the errors. On two years of
		production logs from the MareNostrum supercomputer, our method reduces
		lost compute time by 54\% compared with no mitigation and is just 6\% below the optimal Oracle method. All
		source code is open source.
	\end{abstract}

	
	

	
	\maketitle

	\section{Introduction}
	\label{sec:Introduction}
	
\looseness -1 System resilience is an important requirement for large-scale
clusters, especially in high-performance computing~(HPC), where a single job may
execute for days on thousands of nodes. If any node fails, the job is
terminated, wasting all CPU--hours since the last checkpoint.  One of
the principal causes of hardware failure in HPC clusters is an uncorrected
error~(UE) in main memory~\cite{HP:2016, Giurgiu:Middleware2017,
Schroeder:SIGMETRICS2009, Hwang:ASPLOS2012}. A cost-effective DRAM error mitigation scheme will therefore allow us to maintain reliability as we scale to
larger systems.
 
\looseness -1 The majority of prior work on prediction of corrected~\cite{Costa:ProactiveMemory-ErrorAvoidance, Baseman:DSN17, Sun:FailurePredictionUsingDL, Du:FailurePredictionUsingOnlineLearning, Nie:DSN2018} 
and uncorrected~\cite{Giurgiu:Middleware2017, mukhanov2019workload, wang2021workload} 
memory errors shows great performance in terms of accuracy, precision, and 
recall. It is unclear, however, whether these prediction methods can be used as
the basis for a memory error mitigation scheme that is cost-effective and useful in
practice.  Practical cost-effectiveness and usefulness can only be
evaluated using a cost--benefit analysis, which compares the resources
needed for training, prediction and failure mitigation against the saved
compute time due to successful mitigation~\cite{boixaderas2020cost}.
Saved compute time is quantified in node--hours, which is the sum across
all nodes of the
wallclock time (in hours) that would otherwise have been lost. 
Cost--benefit analysis is complex and
dependent on the failure mitigation strategies and HPC job sizes~\cite{boixaderas2020cost, Das:Doomsday}.
On a given system, the cost of an uncorrected error varies among jobs whose size and duration
can differ by orders of magnitude~\cite{wolter2006s, hart2011deep, yang2013integrating, rodrigo2015hpc, rodrigo2016towards, rodrigo2018towards, simakov2018workload, fan2021deep}, and even within a single job depending
on the time since it started or last performed a checkpoint. 

\looseness -1 
To the best of our knowledge, this paper develops and evaluates the first 
adaptive AI method that decides to trigger the mitigation action depending on both
the likelihood of an error and its potential cost to the current jobs.
Our method is based on reinforcement learning~(RL), and it takes account of
preceding warnings, corrected and uncorrected errors, and node-level events
such as reboots. The RL agent decides when to take active measures to mitigate a potential uncorrected error. 
It is independent of the specific method used for mitigation, so it can be applied to
control various approaches such as node cloning, live job migration or checkpointing.
The only user-defined parameters are the total mitigation cost and whether the job can be restarted from a mitigation point. The method
can therefore be applied to other systems without customization or tuning.
We release all code as open source~\cite{ModelAndSource}.

We train and evaluate the model on MareNostrum~\cite{MareNostrum:guide}, one of
six Tier-0 HPC systems in Europe. At
the time of the study, it comprised 3056 nodes with more than 25,000 memory
DIMMs.  The error logs cover a production period of more than two years, from
October 2014 to November 2016, during which we detected 4.5~million corrected
errors and 333 uncorrected errors.

\looseness -1
The cost--benefit analysis shows a saving of more than 40,000 node--hours over
two years, a 54\% reduction compared with no mitigation. This saving is just
6\% below that of the optimal Oracle prediction scheme.  The evaluation starts
from an untrained model, and is based on time series cross-validation, in order
to avoid overfitting to the fixed historical data~\cite{whiteson2011protecting}.

\looseness -1 We increase confidence in the generality of the method by showing
that it works well for all three major DRAM manufacturers. Evaluation with
different mitigation costs indicates that the model could be applicable to
different failure mitigation approaches.  Finally, the model outperforms other
predictors when the job sizes are scaled up to ten times smaller or larger than
during MareNostrum production.  Applying our method to larger systems would
therefore lead to roughly proportional savings that are larger by orders of
magnitude.

\looseness -1 The rest of the paper is structured as follows.
Section~\ref{sec:Environment} describes the environment and collection of the error and
job logs. Section~\ref{sec:Error-prediction} explains the
Markov decision process formulation of
the problem and its solution using a dueling double deep Q-network. Section~\ref{sec:Evaluation-methodology}
explains the evaluation methodology and Section~\ref{sec:Results} gives the results. Section~\ref{sec:Related-work} is 
the related work and Section~\ref{sec:Conclusions} concludes the paper.

	\section{Environment description}
	\label{sec:Environment}
\subsection{MareNostrum~3 error logs}
\label{sec:error-logs}

Our algorithm is trained and evaluated using memory error and job
logs from two generations of the MareNostrum supercomputer. The error logs were
obtained from MareNostrum~3~\cite{MareNostrum:guide} over more than two years of
production from October~2014 to November~2016.  
At the time, MareNostrum~3 was one of six Tier-0 (largest) HPC systems in
the partnership for advanced computing in Europe (PRACE)~\cite{website:prace}.
It comprised 3056 compute nodes, each with two eight-core Intel Sandy Bridge-EP
E5-2670 sockets and a clock frequency of 2.6~GHz. 
We use the error logs from the compute nodes, excluding the login and test
nodes that are not part of the monitoring infrastructure and whose failures do
not affect the compute jobs.
The jobs executed on MareNostrum~3 were mainly large-scale scientific HPC applications,
and system utilization was generally above 95\%.
During the observation period, we collected data from more than 25,000
\mbox{DDR3-1600 DIMMs}. 
We analyze DIMMs from all three major memory manufacturers, which
have been anonymized and are
referred to as \textit{Manufacturer~A},~\textit{B} and~\textit{C}.  There are
6694, 5207 and 13,419~DIMMs from \textit{Manufacturer~A},~\textit{B}
and~\textit{C}, respectively.

MareNostrum~3 employed single device data correction~(SDDC) ECC.
%
The ECC check is performed on each application memory request and by a patrol
scrubber which periodically traverses physical memory and performs an ECC
check on each location.    

\subsubsection{Corrected errors~(CEs)}

\looseness -1 A daemon, based on {\small\textsf{mcelog}} from
Linux~\cite{Kleen:LK2010}, periodically extracted information related to
corrected errors from the Intel CPU machine check architecture~(MCA)
registers~\cite{Kleen:LK2010}. Each CE was recorded in the log
file, specifying the time stamp, node id, DIMM id, and physical location of
the error including DIMM rank, bank, row and column.\footnote{The mapping from
address to physical location is sensitive manufacturer information, and was
obtained using help from a memory manufacturer.} The log entry also indicates
whether the CE was found by an application memory read or the patrol
scrubber. If there were more than one error within the measured time period,
the MCA registers record the number of errors and provide detailed information
for only one of the errors.  Our logs therefore give the precise number of
CE and they provide detailed error information for a subset
of the errors. We selected a time period of 100\,ms for the daemon, as this was
the shortest time period with a negligible performance overhead.  A shorter
period would increase the size of the sample of detailed error information, but
it would also increase the overhead.  Previous studies perform readings
at a similar~\cite{boixaderas2020cost,Sridharan:SC2012, Sridharan:SC2013,
Sridharan:ASPLOS2015} or larger time period, up to once per hour~\cite{Li:USENIX2010}.  

\subsubsection{Uncorrected errors~(UEs)}

The IBM firmware~\cite{IBM:iDataPlex:2014}, which is part of the MareNostrum~3
monitoring software, logged uncorrected errors, specifying, for each error,
which DIMM failed and whether the UE occurred during an application memory read
or it was found by the patrol scrubber. The log also contains critical
over-temperature conditions, which similarly cause the node to be shut down,
so are counted as equivalent to uncorrected errors. 
It additionally records UE warnings, generated when
the correctable ECC logging limit has been reached or the memory modules are
throttled to prevent an over-temperature condition. UE warnings are input features
to the algorithm but not counted as UEs.

\subsubsection{UE reduction}
\label{sec:UEreductions}
As is the case for many failure events~\cite{Schroeder:TDSC2010, Gupta:SC2017},
uncorrected errors tend to appear in
bursts~\cite{Zivanovic:MEMSYS2019,boixaderas2020cost}.  Burstiness is important
to consider in any study related to UE prediction or mitigation, especially
since, for our dataset at least, the second and subsequent UEs within a burst,
which have no effect on system reliability, are much easier to predict than the
first UE in a burst.  In MareNostrum, whenever a node encountered a UE, it was
removed from production and tested for one week.  This means that only the
first UEs on a node, within a period of one week, have an impact on a production
workload. Filtering the dataset to contain only the first UE in
each burst (of up to a week), reduced the number of UEs from 333~UEs to 67~UEs,
making a major difference to our method's design and evaluation.

\subsubsection{DIMM retirement bias}
\label{sec:retirement_bias}

MareNostrum includes a pre-failure alert, which identifies DIMMs that are close
to failure. Such DIMMs were retired in order to reduce the incidence of
uncorrected errors in the production system.
Over the two-year production period, 51~DIMMs were retired by the system
administrators. This action is recorded in the system log with the date and time. 
We could not determine the specific reasons for DIMM retirement.  Surprisingly,
most of the retired DIMMs experienced no preceding corrected or uncorrected
errors in the error log, and they performed no node boots in the days before
the retirement. A recent IBM study~\cite{Giurgiu:Middleware2017} mentions hundreds of sensors that are used by the system
integrators to predict component failures. We had no access to these sensors in
the system under study.

\looseness -1
Preventive DIMM retirement
introduces a bias in training and evaluation that we were unable to
avoid. Since it is impossible to know whether an event followed by DIMM
retirement would otherwise have been followed by an uncorrected error,
we remove all such samples from training and evaluation.

\subsubsection{Quantitative analysis}

Zivanovic et al.~\cite{Zivanovic:MEMSYS2019} perform detailed analysis of the
same MareNostrum~3 error logs that are used in our study. They apply several
different methods for quantitative and statistical analysis of the DRAM
corrected and uncorrected errors, as well as memory system faults.
The authors also compare the results for different DRAM manufacturers and DRAM cell technologies.
The number of UEs in Zivanovic et al.\ differs from our results
by about 6\%. This is
because they exclude DIMM critical over-temperature conditions and ignore address DIMM
retirement bias (Section~\ref{sec:retirement_bias}). Their analysis is a
complement to this paper and it gives confidence that the error logs used for
our study are representative of typical DRAM failures.

\subsection{MareNostrum~4 job logs}
\label{sec:mn4-jobs}

\looseness -1 The log of the jobs was obtained from the general-purpose block of the
successor, MareNostrum~4~\cite{MareNostrum4}, which has 3456 nodes, each
with two 24-core Intel Xeon Platinum sockets at 2.1~GHz.  The job log covers
the production period from Mar.~2018 to Mar.~2019. For practical reasons, it
was impossible to use error and job logs from the same period and
system.  We believe that combining logs from different machines and
different production time periods does not significantly change any of the
conclusions.
The process of combining error logs and job logs from both systems is elaborated upon in Section~\ref{sec:training}.
The only major aspect that could not be addressed in this study is
any possible correlation between the occurrence of errors and recorded job
characteristics such as wallclock duration and number of nodes.

\looseness -1 The job log was collected by Slurm~\cite{yoo2003slurm}, which is
the job scheduler used to run jobs on MareNostrum~4. We extracted the log 
using the {\small\textsf{sacct}} command, which provides the job's submission
time, start and end times, allocated node IDs, and other information.

	\section{Adaptive Error Mitigation Control}
	\label{sec:Error-prediction}

\subsection{Background}
\label{sec:background}

Reinforcement learning (RL) is one of the most promising machine learning (ML)
approaches for tackling hard control problems. The main difference between RL
and other ML methods is that RL learns in a dynamic environment. An RL approach
is composed of three main elements: the \emph{agent}, the \emph{environment},
and the \emph{reward function}.  The agent interacts with the environment,
which is a model of the system, by observing the current state of the
environment and taking a specific action. In response, the environment changes the
state and the agent receives a reward.  The agent's goal is to discover
and take the optimal sequence of actions in the environment in order to
maximize the cumulative sum of the rewards. The learned mapping between the
states of the environment and the actions is known as the policy.  The agent
learns, i.e. it updates its policy, by exploring and interacting with the
environment.

An RL problem is usually formalized as a Markov decision process
(MDP)~\cite{bellman1957markovian}, which has four elements: 1) the set of
states of the environment, 2) the set of possible actions, which may depend on
the current state, 3) the probabilities of moving among states depending on
the action, and 4) the rewards associated with these transitions. Each
will be explained in the context of DRAM error mitigation in
Section~\ref{ssec:adaptiveErrorMitigation}.

The goal of an RL algorithm associated with an MDP is to find a policy that maximizes the discounted sum of the rewards from each time-step:
\begin{equation}
	\sum_t \gamma^t R(t),
	\label{eqn:value}
\end{equation}
where $\gamma$ (gamma) is the discount factor and $R(t)$ is the reward obtained at time $t$. The discount factor is between 0 and 1, and its value controls the tradeoff between taking an immediate high reward and maximizing rewards over time.

\looseness -1 Q-learning~\cite{watkins1989learning} is an RL algorithm in which the agent
independently learns the value of each action in each state. The value
is the expected accumulated reward, from Equation~\ref{eqn:value}, of
taking a given action in the starting state and following the learned policy thereafter.
All these values, for every possible state and action, are stored in a table known as the
Q-function.

Many problems, including DRAM error mitigation, have states with a large number
of dimensions, some of which are continuous or have so many values they are
effectively continuous. Learning and storing the value of each state
individually would be prohibitive in terms of training time and storage, so
various approaches are used to approximate the Q-function. One of the best
known approaches, deep Q-learning~\cite{mnih2013playing}, approximates
the Q-function using a deep neural network known as a deep Q-network. The
agent trains the network to minimize a loss function, which quantifies the error
introduced by approximating the Q-function.

We employ two known approaches: dueling double deep
Q-network~(DDDQN)~\cite{wang2016dueling} and prioritized experience
replay~(PER)~\cite{schaul2015prioritized}. A DDDQN is both a double deep Q-network and a dueling network
architecture. A double deep Q-network uses two
different neural networks, one to select the action and the other to evaluate
it, mitigating a well known overestimation bias due to self-evaluation. A dueling network architecture splits
the Q-function into two parts, known as the value function and the advantage
function. The value function is the expected reward, according to the policy,
in a given state and the advantage function indicates how much better each
action is compared with the expected reward. This approach is known to converge more
rapidly to an optimal policy~\cite{wang2016dueling}.
PER
stores the outcomes of past experience in the environment, and training is done
based not only on the current actions but mini-batches of past experiences. 
The experiences with higher discrepancies between predicted and actual outcomes are considered more important, guiding the prioritization mechanism so as to improve the efficiency of
learning.

\subsection{MDP formulation of UE mitigation control}

\label{ssec:adaptiveErrorMitigation}

\subsubsection{State and features}

\looseness -1 The features in the state are listed in Table~\ref{tab:Features}.
The CE, UE and system state features, i.e.\ all features except the potential
UE cost, are derived from the error log events observed in the nodes of
MareNostrum~3.  All features are calculated at each time step, and provided
directly to the agent.  In addition, for the two features annotated with an
asterisk, the feature variation over time is calculated as: 
\begin{equation}
	\small
	\textit{Feat. variation ($\Delta t$)} = \dfrac{\textit{Feat. value (Prediction moment)}}{\textit{Feat. value (Prediction moment - $\Delta t$)}}, 
	\normalsize
	\label{eqn:feature-variation}
\end{equation}

\begin{table}[tbp] 
	\caption{Features used for UE mitigation control}
	\begin{center}
		\begin{tabular}{@{}l@{}}
			\toprule
			\textbf{Feature in state (per node)} \\
			\midrule
			\textit{Corrected errors~(CEs):} \\
			\quad Number of corrected errors since the last event\\
			\quad Number of CEs since the beginning of operation$^{\mathrm{*}}$ \\
			\quad Number of ranks, banks, columns and rows with CEs \\
			\quad Number of DIMMs with CEs \\[0.5mm]
			\textit{Uncorrected errors:} \\
			\quad Number of UE warnings since the beginning of operation\\[0.5mm]
			\textit{System state:} \\
			\quad Time since the last node boot (start) \\
			\quad Number of node boots$^{\mathrm{*}}$ \\
			\textit{Workload:} \\
			\quad Potential uncorrected error~(UE) cost \\[0.5mm]
			\bottomrule\addlinespace[1.5mm] 
			\parbox{\columnwidth}{\small $^{\mathrm{*}}$~Feature variation over time (Equation~\ref{eqn:feature-variation}) is calculated for this feature.} \\
		\end{tabular}
		\label{tab:Features}
	\end{center}
\end{table}

\noindent where $\Delta t$ is the time increment. The feature variation over
time is calculated for $\Delta t$ equal to 1 minute and 1 hour, and it is set
to zero if the denominator in the above equation is zero.

The potential UE cost depends on the workload, and is the total
number of node--hours that would have been lost if a UE had occurred at the
moment the agent is invoked:
\begin{equation}
		\mathit{UE\_cost} = \mathit{nodes} \times \mathit{potential\_lost\_wallclock\_time}.
		\label{eqn:ue-cost}
\end{equation}

In case the mitigation allows job restart, e.g. checkpointing, then \textit{potential\_lost\_wallclock\_time} is the time since the beginning of the job or the last mitigation. Otherwise it is the time since the job started.

\subsubsection{Actions}

There are always two actions available to the agent: it can either request a job
mitigation (action $a$ equals 1) or do nothing ($a$ equals 0).

\subsubsection{State transitions}

\looseness -1 After taking an action, the agent has nothing to do until the
next event. If the next event is a UE, then the whole node is shut down and the
job is terminated without invoking the agent.
Otherwise, the state will transition to correspond to the
next event.  When the events are taken from a historical log, the CE, UE and
system state features and their relative rates of change over time do not
depend on the agent's last action. The potential UE cost, however, always
depends on whether a mitigation was performed. If the agent did not
trigger a mitigation, 
then the potential UE cost is increased by the job's elapsed
node--hours. If the agent requested a mitigation, then the potential UE cost is first set
to zero to reflect the mitigation action, and then it is increased by the job's
elapsed node--hours.  There is a minimum wallclock time between state
transitions of one minute, so that events occurring within the same minute are
combined.  For our system we found that a higher frequency than once per minute
would increase the overhead but make no improvement to the cost--benefit
analysis.

\subsubsection{Reward}
\label{ssec:reward}

The reward function is calculated based on the lost node--hours of the system:
\begin{equation}
	R_a = - a \times \textit{mitigation\_cost} - \textit{UE\_occurred} \times \textit{UE\_cost},
	\label{eqn:reward}
\end{equation}

\noindent where $a$ is the action (1 for a mitigation, 0 otherwise) and \textit{UE\_occurred} equals
1 if a UE occurred following the action and 0 otherwise.

\subsubsection{Error mitigation actions and cost} 

The RL agent decides when to perform mitigation actions, based on the
error-related features and the potential cost of an error. If the agent
requests a mitigation, then the actual error mitigation is performed by the
\emph{environment}. As such, the agent is independent of the specific
mitigation method, and the only mitigation-related parameter is the mitigation cost.

The training and evaluation in Section~\ref{sec:Results} uses a error mitigation
cost of 2 node--minutes, following estimations from a recent study of
Das et al.~\cite{Das:Doomsday}. 
The study analyzes various actions that can
mitigate the impact of node failures, such as live job migration, node cloning
and checkpointing, and concludes that 2\,min suffice for most of these actions.
We also consider the mitigation cost of 5 and 10 node–minutes, which is the
checkpointing time considered in various previous studies~\cite{oliner2005performance, iskra2008zoid, ross2006parallel, elliott2012combining, elnozahy2009system, bergman2008exascale, frank2021improving}.

Finally, the UE cost is calculated using
Equation~\ref{eqn:ue-cost} with the timestamp of the UE.  This means that the
UE cost always includes the full time elapsed between the last mitigation and
the actual UE.  The goal of the agent is to maximize the cumulative discounted
sum of these negative costs.

\subsection{Solving the Markov Decision Process~(MDP)}

\begin{figure}
	\centering
	\includegraphics[width=\columnwidth]{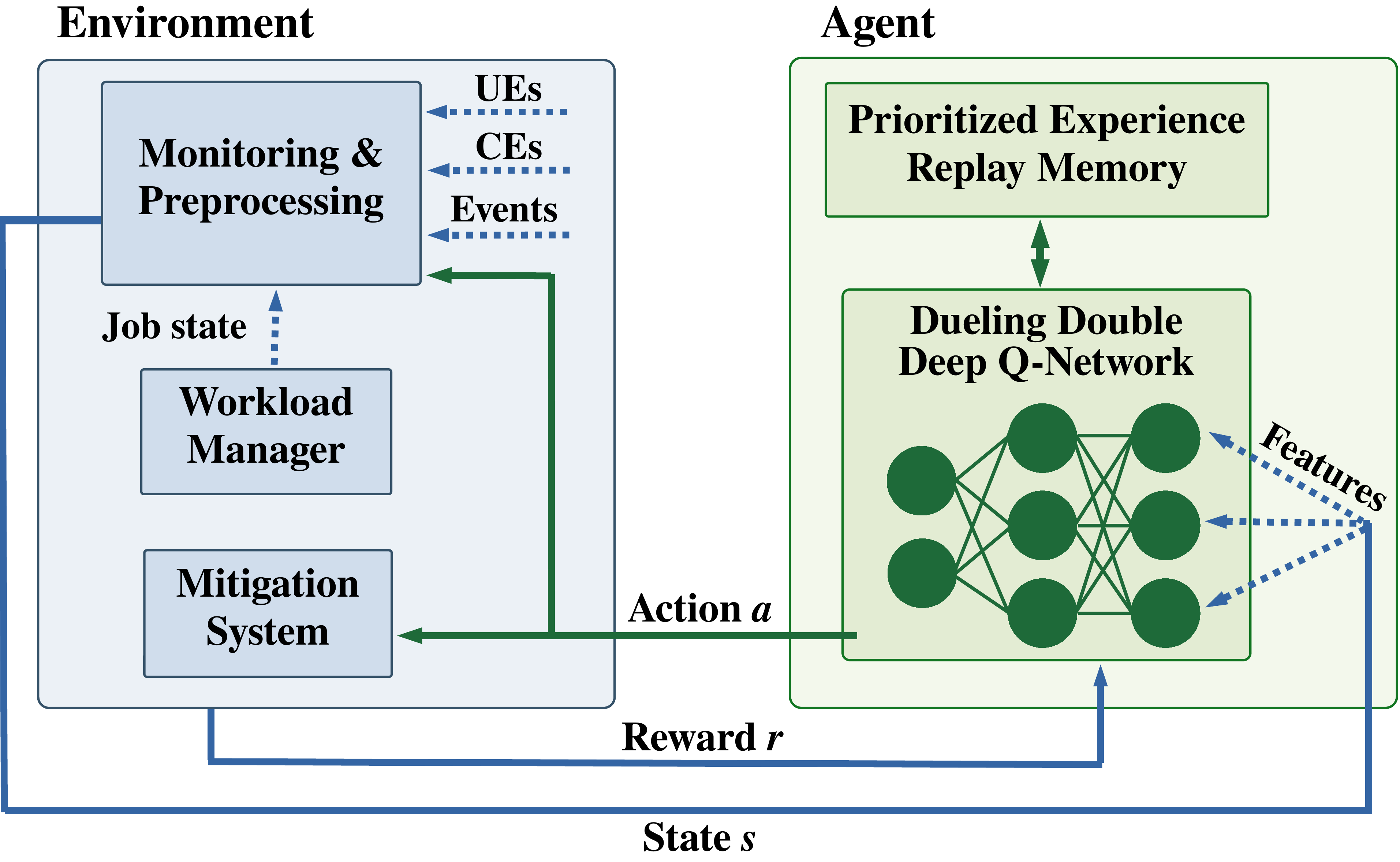}
	\caption{Interaction between RL agent and environment for adaptive UE mitigation.}
	\label{fig:agent-and-environment}
\end{figure}

\subsubsection{Overall approach}
\looseness -1
Figure~\ref{fig:agent-and-environment} illustrates how the RL agent solves the
MDP problem described in the previous section. The diagram shows the
environment on the left and the agent on the right.  In our implementation
based on historical logs, the environment obtains the UEs, CEs, events and job
state from the logs, as described in Section~\ref{sec:Environment}. In a real
system, the environment would collect this information from the
monitoring daemons and workload manager. The environment passes
the state to the agent, which supplies the state features to the neural network 
to determine the action. In order to avoid overestimation of the action values
and to learn a better policy, we use
a dueling double deep Q-learning neural network. Given the
action selected by the agent, the environment either performs a mitigation or
it does not.  On the next event, it calculates the reward for the last action,
depending on the last time that mitigation was performed and whether or not the next event is a
UE.

\subsubsection{Architecture}

As an approximator for the Q-function, the agent uses a deep neural network composed of the input features, four hidden layers with 256, 256, 128, and 64 neurons, respectively, and a single output to indicate whether or not to mitigate.

\subsubsection{Training}
\label{sec:training}

\looseness -1
Training is divided into episodes, each of which is a ``run'' of the
agent in its environment from an initial to a terminal state. In
our context, a single node is chosen randomly and the episode runs from the
beginning to the end of the split (Section~\ref{ssec:time-series-xval}), taking all events
on that node.  A sequence of jobs is randomly chosen to run
on the node. The jobs are weighted by the number of nodes on which they execute,
in order to maintain the correct job distribution.
During training, the
agent learns to maximize the reward by improving its future actions 
based on its current experience.

\subsubsection{Class imbalance}
\looseness -1 During the two-year period, there were 67 UEs (following UE reduction of
Section~\ref{sec:UEreductions}) out of a total of 259,270 events (after merging events in the same minute).
This imbalance of 3.5 orders of magnitude between the numbers of UEs and events
causes the learning process to be slow. We therefore use a form of experience replay
known as prioritized experience replay~(PER)~\cite{schaul2015prioritized} (Section~\ref{sec:background}),
which speeds up learning by prioritizing experiences that are expected to
result in more learning progress. In our case, PER demonstrated effectiveness in handling class imbalance without the need for additional techniques commonly employed in ML for this purpose.

	\section{Evaluation methodology}
	\label{sec:Evaluation-methodology}

\subsection{Time series nested cross-validation}
\label{ssec:time-series-xval}

\begin{figure}
	\centering
	\includegraphics[width=\columnwidth]{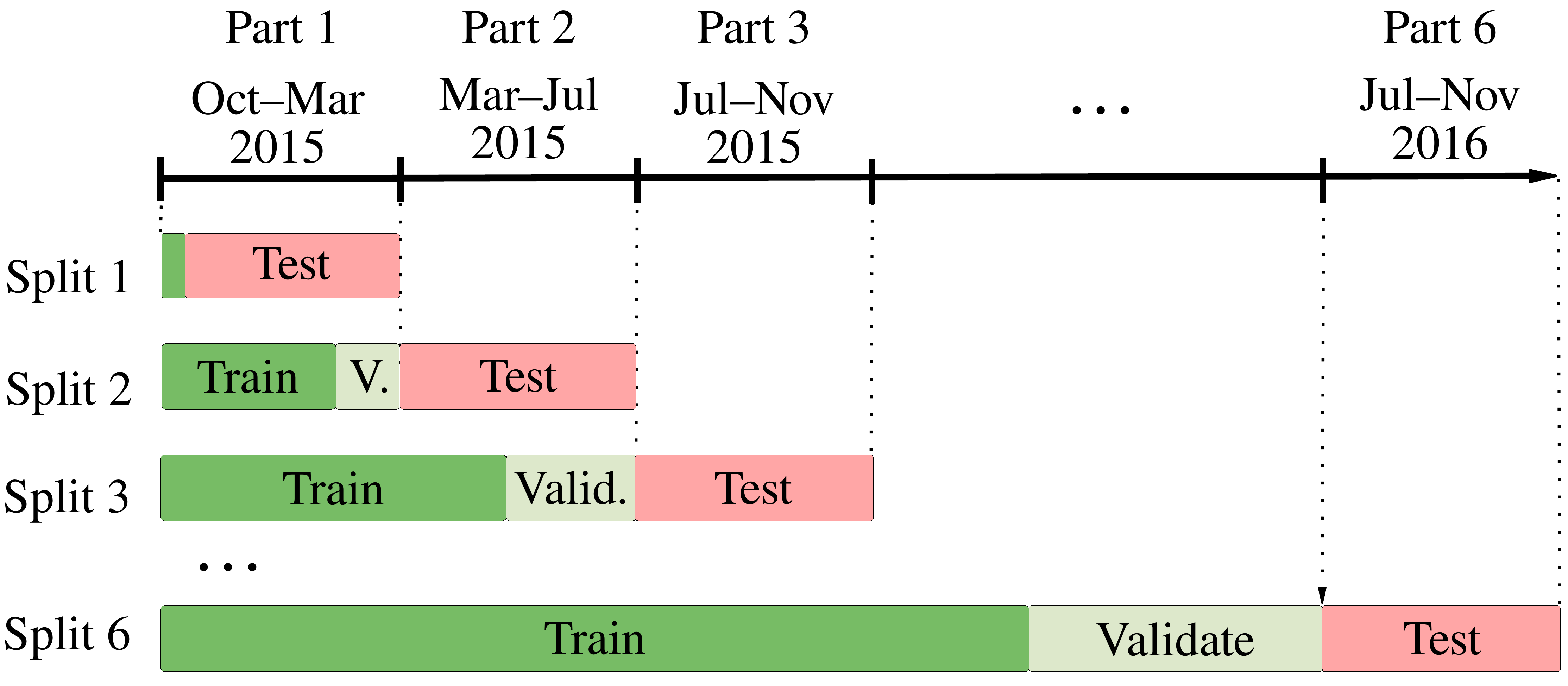}
	\caption{\looseness -1 Evaluation using time series nested cross-validation.
		The error log is divided into six equal parts and evaluation for
		each split is divided into training (with multiple hyperparame-
		ters), validation (to find the best hyperparameters), and testing.}
	\label{fig:nested-xval}
\end{figure}

We evaluate our RL approach on production data using time series nested
cross-validation, which is a well-known technique for determining how well a
model performs and generalizes in a setup similar to how it would be used in
practice. Figure~\ref{fig:nested-xval} summarizes this methodology.  The error
log is divided into six equal parts (shown horizontally), each of which
corresponds to roughly four months of data. These parts are used to create six
splits (shown vertically), each of which allows a separate evaluation on a part
of the test data, using a model trained with hyperparameters chosen using
data that precedes that part. The evaluation for each split is divided into
training (with multiple hyperparameters), validation (to find the best
hyperparameters), and testing (to evaluate the cost--benefit). Although it
is common in RL to only present the results for an agent using the best
hyperparameters, our use of historical data rather than an interactive setting
means that this approach would introduce bias~\cite{whiteson2011protecting}.

In the hyperparameter tuning phase we adjust the learning rate of the neural
networks, the discount factor $\gamma$, the update and synchronization
frequencies of the two networks and some of the parameters of the prioritized
experience replay, such as the batch sample size. We perform a first round of
random search with 60 sets of hyperparameters, selecting the hyperparameters of
the best performing agent on the training data and running another round with a
narrowed search space, close to those best hyperparameters, finally selecting
the best performing agent on the validation set. Each agent is trained with
20,000 episodes,\footnote{Although the error and job logs are fixed historical
records, each episode is different because it uses a random sequence of jobs
(Section~\ref{sec:training}).}
which proved to be sufficient to achieve stable decisions with good rewards.

\looseness -1 Each split, except the first, allocates the first 75\% of the time before the
new part for training and the remaining 25\% for validation.  The training set
is used to train multiple agents with different hyperparameters, starting from
an untrained model for the first split and a mix of previously trained and
untrained models for each subsequent split.  The validation set is then used to
select the best performing agent. If there are no UEs in the validation set,
which happens when partitioning by manufacturer due to the low frequency of UEs, then we
select the best performing agent for the training set. This introduces some
bias but it avoids the risk of choosing a weak mitigation policy. The best
agent is then evaluated using the testing dataset. The testing dataset is then
incorporated into the training and validation datasets of the next part, which
is again divided in the same ratio of 75\% to 25\%.  The first split is
slightly different, because it employs the first two weeks from the log for
both training and validation, and the rest of the first three-month part is
used for testing. This approach allows us to include almost all of the
production log in the evaluation.  The overall result is the total node--hours
lost, due to mitigations and UEs, accumulated across all six splits used for
evaluation.

\subsection{Approaches under evaluation}
\label{ssec:approaches}

We evaluate six prediction-based approaches 
that can be used to decide when to perform UE mitigation:

\begin{itemize}
	\item \textit{Never-mitigate} does not initiate any UE mitigation. It leads to the maximum possible UE cost but the minimum possible mitigation cost.
	\item \textit{Always-mitigate} triggers a UE mitigation for
	every event in the error log.  It has the minimum possible UE cost but the
	maximum possible mitigation cost, among policies that signal mitigations only
	when there is an error event. It is implicitly a form of predictor, since an
	event in the error log is treated as an indicator of an upcoming UE.
	\item \looseness -1 \textit{SC20-RF} is the state-of-the-art random forest~(RF) predictor of Boixaderas et
	al.~\cite{boixaderas2020cost}. The authors applied six machine learning
	classifiers and found that random
	forest, with random under-sampling to address class imbalance, provided the
	best results. The output of the random forest predictor is a value from 0 to 1
	that represents the probability of an uncorrected error. A mitigation is
	triggered if the value exceeds an externally provided threshold
	parameter.  We start from an untrained model, but provide
	maximum advantage to SC20-RF by using the optimal threshold parameter.
	\item \looseness -1 \textit{SC20-RF-2\%} and \text{SC20-RF-5\%} are the SC20-RF policy with
	realistic (suboptimal) values of the threshold parameter, differing from the optimal value
	by 2\% and 5\% respectively.
	\item \textit{Myopic-RF} is an extension of \textit{SC20-RF} that adapts to
	the current potential UE cost. It triggers a mitigation action if the
	expected cost from a UE (probability of a UE multiplied by the cost it
	would have), without mitigation, is greater than the cost of
	mitigation.  The probability of a UE is estimated by the RF predictor.
	\item \textit{RL} is the reinforcement learning approach presented in this paper, always starting from an untrained model.
	\item \textit{Oracle} signals a UE mitigation on the last event before each UE. It therefore
	performs the minimum number of mitigations necessary to predict the maximum
	number of UEs. It is the optimal strategy assuming that all mitigations are performed following
	events in the log.  It is not a realistic policy for implementation, but it
	allows us to quantify the room for improvement.
\end{itemize}

\subsection{Total costs in node--hours}
\label{ssec:total-costs}

\looseness -1 All cost--benefit calculations show the total number of lost node--hours,
i.e.\ the sum of the UE cost and mitigation cost. The UE cost is
computed using Equation~\ref{eqn:ue-cost}, calculated at the precise time
of each UE. The mitigation cost is the total cost of the mitigation actions plus,
for SC20-RF, Myopic-RF and RL, the cost of all training and validation used to create
the model. The time to determine the optimal threshold
parameter for SC20-RF is not included in the evaluation.

There is a 7\% difference between our results for SC20-RF in
Section~\ref{sec:Results} and the results in its original
publication~\cite{boixaderas2020cost}. This is because the previous study
does not compute the lost compute time between the mitigation action and a
UE, when a UE occurs inside the prediction window (period for which the
prediction is made). Our results always count the full UE cost, including the
full time period between the last mitigation action and the UE.

\subsection{Classical machine learning metrics}
\label{ssec:classical-metrics}

\looseness -1 Previous studies that propose error prediction methods
\cite{Giurgiu:Middleware2017, Sun:FailurePredictionUsingDL,
Costa:ProactiveMemory-ErrorAvoidance, Du:FailurePredictionUsingOnlineLearning,
boixaderas2020cost, wang2021workload} are evaluated using standard 
prediction metrics such as recall or precision. For completeness,
and to allow a direct comparison with these methods, we also evaluate our RL
method using these standard metrics.

In order to perform this evaluation, we classify the actions of the prediction-based methods as:

\begin{itemize}
	\item \textbf{True positives (TPs)}: Number of UEs that were mitigated. 
	\item \textbf{False negatives (FNs)}: Number of UEs that were not mitigated. 
	\item \textbf{False positives (FPs)}: Number of mitigations minus the number of TPs.
	\item \textbf{True negatives (TNs)}: Number of non-mitigations minus the number of FNs.
\end{itemize}

Following Boixaderas et al.~\cite{boixaderas2020cost}, the
classical machine learning metrics assume a prediction window of 1 day. This
means that a UE is counted as successfully mitigated, i.e.\ as a true
positive, if at least one mitigation action
completed within the previous 24\,hours, i.e. was initiated within the previous
24\,hours minus the 2 node--minutes mitigation overhead specified in
Section~\ref{ssec:adaptiveErrorMitigation}. The remaining UEs are counted as
not mitigated, i.e. as false negatives. We only employ the prediction
window to calculate the classical machine learning metrics, which need a
binary classification into mitigated or not mitigated. The cost--benefit
calculation uses the real UE and mitigation costs described in
Section~\ref{ssec:total-costs}.

The number of mitigations is the number of times that the policy selects the
mitigation action ($a=1$ in the case of RL). A single UE may be mitigated
multiple times within the 24-hour period, but only one of these mitigations
(which can be imagined to be the one that happens last) is a true positive. The
other mitigations are redundant and counted as false positives.  The number of
non-mitigations is the number of times that the policy selects not to mitigate
($a=0$ for the RL agent) plus the number of UEs that have no event in the
preceding time window of 1 day. If there is no event within the 24\,hours
before the UE, then none of the policies, all of which mitigate only in
response to an event, has an opportunity to mitigate the UE. Nevertheless,
since the UE was not mitigated, it must be counted as a false negative, to
avoid biasing our results by ignoring the hardest-to-mitigate UEs. We avoid
this bias by assuming that the system makes an implicit ``no-mitigate'' false
negative action for these UEs.

\textbf{Recall} is the proportion of actual positives that are correctly identified as such. In our case, this metric refers to the fraction of UEs that are correctly predicted:

\begin{equation*}
\textit{Recall} = \dfrac{\textit{Correctly predicted UEs}}{\textit{Total UEs occurred}}  = \dfrac{\textit{TPs}}{\textit{TPs + FNs}} 													  
\end{equation*} 

\textbf{Precision} refers to the percentage of observations
classified as positives that are true positives.  In our case, the precision
refers to the ratio between correctly predicted/mitigated UEs and the
total number of mitigations performed:
\begin{equation*}
	\textit{Precision} = \dfrac{\textit{Correctly predicted UEs}}{\textit{Total mitigations}} = \dfrac{\textit{TPs}}{\textit{TPs + FPs}} 
\end{equation*}

\subsection{Generality to other job sizes}
\label{ssec:dram-manufacturer}

\looseness -1 We increase the confidence
in the generality of our method to different HPC system architectures and HPC job sizes.
To consider different architectures, we partition the MareNostrum~3 error logs
by DRAM manufacturer, generating 
three smaller subsystems,
of size 6694, 5207 and 13,419 DIMMs,
from anonymized Manufacturers A, B and C respectively.  With few
exceptions, all DIMMs in a given node are from the same DRAM manufacturer.
Firstly, we trained and evaluated the method on the whole system,
\textit{MN/All}.  Secondly, we performed separate training and testing for each
subsystem comprising a single DIMM manufacturer: \textit{MN/A}, \textit{MN/B}
and \textit{MN/C}.  Finally, we give results for the sum of the three
subsystems, \textit{MN/ABC}, which differs from \textit{MN/All} only because it
uses three separately trained models.

\looseness -1 To consider different HPC
job sizes, we perform a job size sensitivity analysis. This is necessary because HPC jobs are known to differ in size and duration by orders of magnitude~\cite{wolter2006s, hart2011deep, yang2013integrating, rodrigo2015hpc, rodrigo2016towards, rodrigo2018towards, simakov2018workload, fan2021deep}. We investigate the effect on the
cost--benefit calculation of job sizes up to ten times smaller or ten times
larger than those seen on MareNostrum~4.
Future work could consider how the results would differ for cloud platforms,
using the public datasets from Microsoft Azure~\cite{cortez2017resource} or
Google Borg~\cite{clusterdata:Wilkes2020}, or for other HPC systems using public
logs~\cite{hujilogs} or synthetic trace
generators~\cite{cirne2001comprehensive}.

\subsection{Generality to other hosts and memory architectures}

\looseness -1 Our results are for a single supercomputer and memory system
architecture (with three different vendors). To fully test the generality of
our method, it is important to consider other CPU or GPU host architectures
with different error correction schemes and error logging capabilities. It
would also be interesting to analyze the impact of on-die ECC, as supported by
current HPC DIMMs, which transparently corrects errors in the memory devices.
On-die ECCs are not standardized by JEDEC or as part of the host--memory
interfaces, so the impact could be very different for different manufacturers.
Modern memory interfaces, HBM, DDR DIMMs and memory-over-CXL enable more
heterogeneity in the memory system, which may also impact the results.

Our RL method is released as open source. Since it has no user-supplied
parameters, the method can be applied without any customization or tuning to
different host and memory system architectures. We encourage the community to
evaluate the method on different systems and share their findings.

	\section{Results}
	\label{sec:Results}

\subsection{Cost--benefit analysis}
\label{sec:results-cost-benefit}

\begin{figure}
	\centering
	\includegraphics[width=\columnwidth]{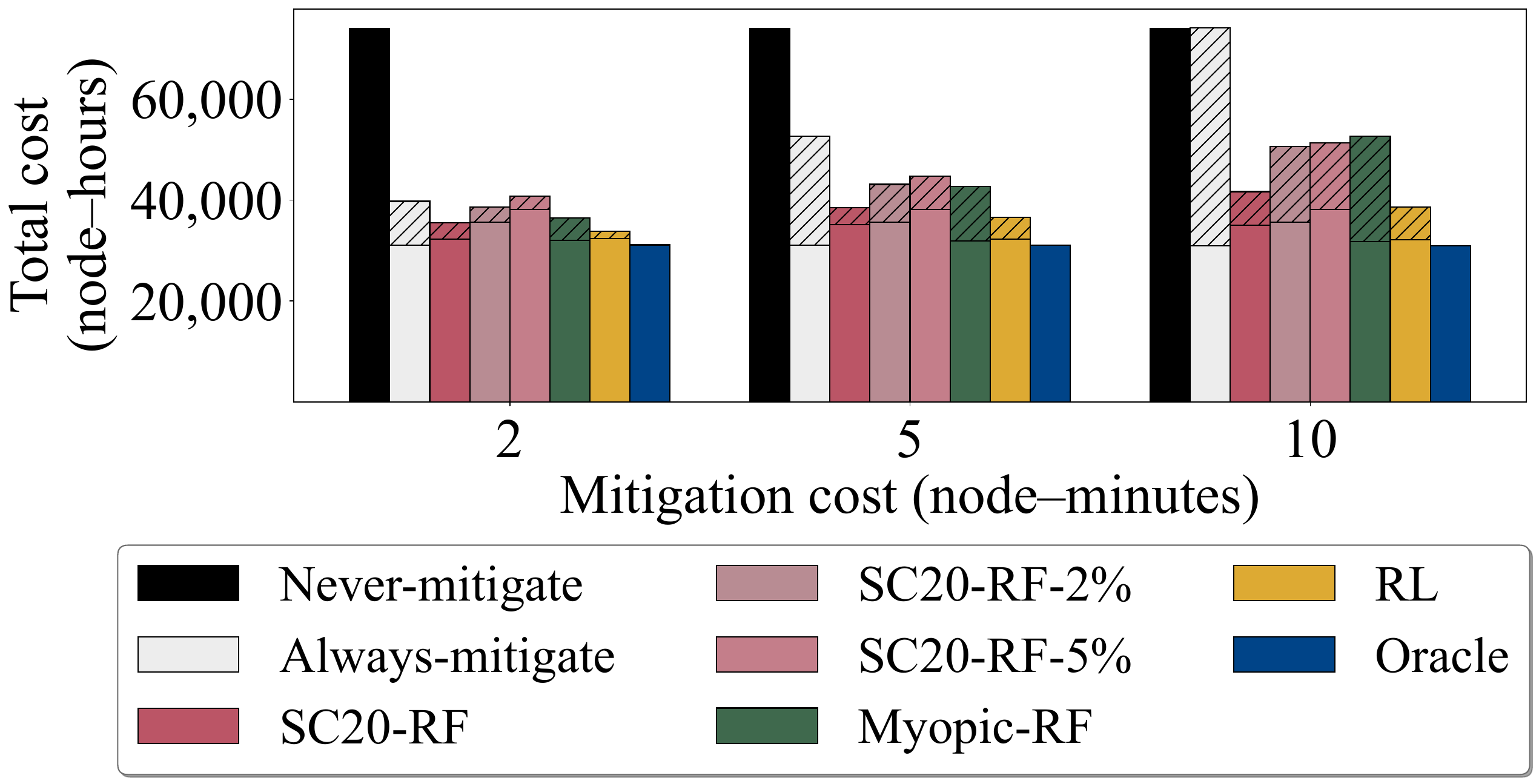}
	\caption{Total cost for MN/All, as the sum of UE cost (solid color) and mitigation cost (with dashes). The RL agent has lower total
		cost than the other approaches due to a much lower mitigation
		cost. Unlike SC20-RF it is not sensitive to a user-supplied
		parameter. The results are stable for mitigation costs between
		2 node–minutes and 10 node–minutes.}
\label{fig:total-costs-mitigation}
\end{figure}

\looseness -1 Figure~\ref{fig:total-costs-mitigation} shows the overall results
of the cost--benefit analysis. The $y$-axis is the total cost over the
two-year production time period, which is the sum of the cost of the UEs~(solid
color) and the cost of mitigations~(with dashes), calculated as the sum across
all six splits in the time series nested cross-validation
(Section~\ref{ssec:time-series-xval}).  The eight bars for each scenario
correspond to the eight approaches described in Section~\ref{ssec:approaches}.
These are the two baseline policies, Never-mitigate and Always-mitigate (on
every error-related event), the state-of-the-art SC20-RF policy with optimal
and suboptimal threshold parameters, Myopic-RF (our adaptive extension of
SC20-RF), the RL approach of this paper, and the optimal Oracle policy. Results are
given for mitigation costs of 2, 5 and 10~node--minutes.

\looseness -1 In all scenarios, the Never-mitigate policy has a large cost,
underscoring the need for some kind of UE mitigation approach. Its cost is
entirely due to UE costs, so it is independent of the mitigation cost, at
74,035 node--hours over the two-year production period. The simplest mitigation
strategy, Always-mitigate, is effective for a small mitigation cost of
2~node--minutes, reducing the cost by 46\% to 39,769 node--hours. It has the
lowest possible UE cost for our dataset and approach triggered by events in the
log, but also the highest mitigation cost, at 8642 node-hours. As the
mitigation cost increases, however, the cost of Always-mitigate increases
dramatically, and for a cost of 10~node--minutes, it is slightly worse than
Never-mitigate (see Section~\ref{ssec:sensitivity} for a sensitivity analysis
using smaller and larger job sizes). Always-Mitigate implicitly includes a form
of prediction, since ``always'' means that any kind of event in the log is
treated as an indicator of an upcoming UE.

The random forest predictor of SC20-RF, with the optimal choice of the
user-defined threshold parameter, reduces the total cost significantly. For a
2~node--minute mitigation cost, it reduces the cost by 52\% compared to
Never-mitigate to 35,543 node--hours. The decision threshold for SC20-RF needs
to be selected carefully, as it significantly affects the machine learning
metrics and cost--benefit analysis~\cite{boixaderas2020cost}. The results
verify this, as with a SC20-RF threshold of just 2\% (or 5\%) from optimal, the
total cost of SC20-RF increases to 38,645 (40,740) node--hours.

The Myopic-RF policy, while seeming to be a reasonable approach, has
consistently worse results than SC20-RF. For a 2~node--minute mitigation cost,
this increase is small, to 36,432 node--hours, but the gap widens considerably
as the mitigation cost increases.  The disappointing results for Myopic-RF
arise because although the output from the RF predictor increases with the
likelihood of error, it is not a reliable probability value, as assumed by
Myopic-RF.

\looseness -1 The RL approach consistently reduces the total cost 
to 54\% below Never-mitigate, at 33,843 node--hours. For a 2~node--minute
mitigation cost, the advance over SC20-RF is mainly due to the lower mitigation
cost, which is approximately 55\% lower than that of SC20-RF. As the mitigation
cost increases, a better choice of when to apply mitigation gives a similar
mitigation cost but somewhat lower total cost. Finally, the Oracle reduces
the total cost by 58\% to 31,129 node--hours. The negligible total mitigation
cost of the Oracle shows that the mitigation cost of the other approaches is
almost entirely unnecessary mitigations (false positives).

\looseness -1 All results include the cost to train and validate the model, where applicable.
The cost of SC20-RF is ``on the order of
node--minutes''~\cite{boixaderas2020cost} and the cost for RL is less than
twenty node--hours per year. The RL agent has a greater training
and validation cost than SC20-RF, but the difference is negligible compared
with the additional saved node--hours. In addition, SC20-RF has a hidden cost
to determine the optimal value of its threshold parameter. This cost is not
quantified in our results, but it could be significant.

In summary, these results show that our RL approach, compared with the state-of-the-art
SC20-RF, reduces the total cost by 5\% and narrows the distance from the
optimal Oracle by more than a third.

\begin{figure}
	\centering
	\includegraphics[width=\columnwidth]{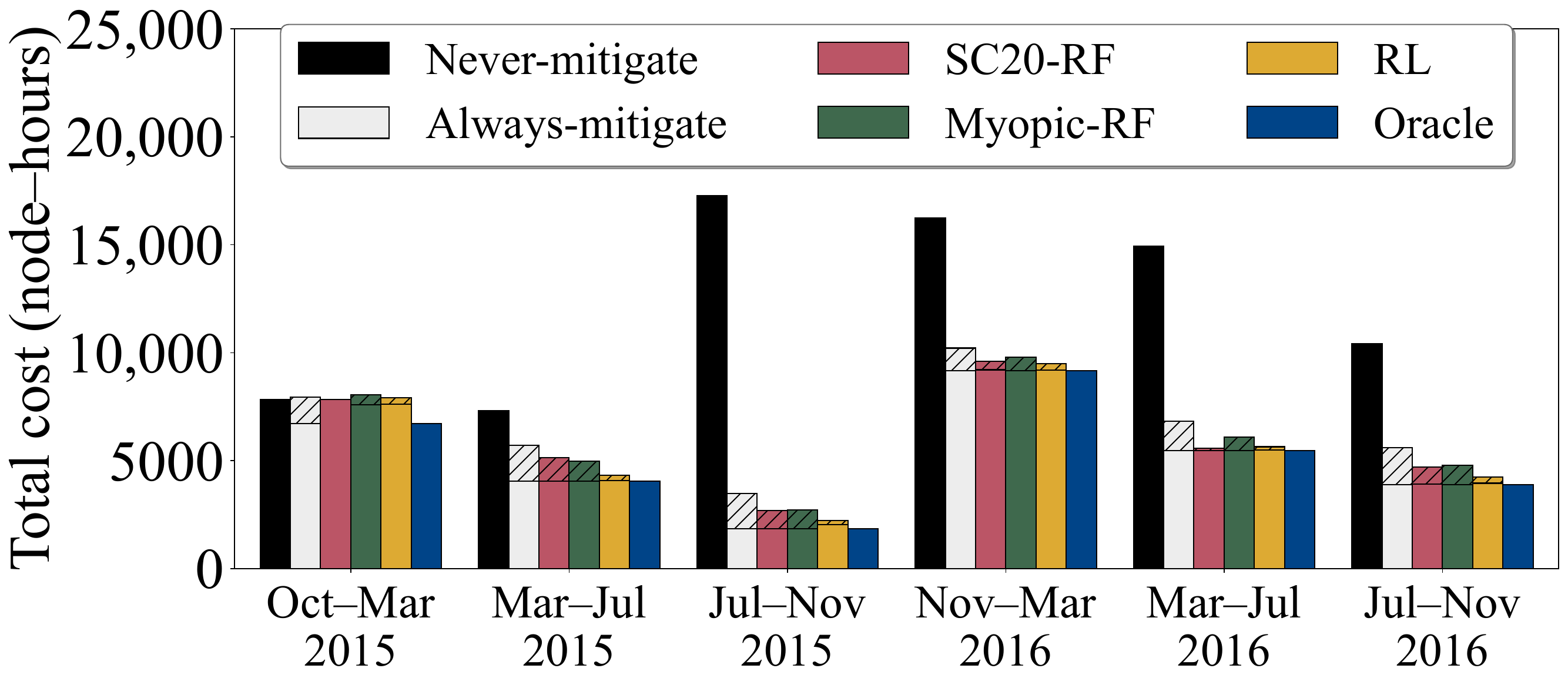}
	\caption{Time series nested cross-validation for MN/All with 2~node--minute mitigation cost, starting
		from untrained models. The
		total cost is the sum of UE cost (solid color) and mitigation cost (with dashes).}
	\label{fig:cross-validation}
\end{figure}

\subsection{Time series nested cross-validation}

Figure~\ref{fig:cross-validation} shows the complete results from the time
series cross-validation for MN/All. The $y$-axis is again the total cost. The
$x$-axis is time, showing a separate set of results for each split, each of
which corresponds to a roughly four-month period during the MareNostrum~3
operation.  The sum across time of the values in
Figure~\ref{fig:cross-validation} match the 2~node--minute bars of
Figure~\ref{fig:total-costs-mitigation}.  We see that the relative performance of all six
approaches is stable over time.  In all six periods except the first,
Never-mitigate consistently has the highest cost, showing a constant need for
some kind of error mitigation throughout operation. SC20-RF outperforms
Always-mitigate in all of the six periods, due to the much lower mitigation
costs for a normally similar UE cost. Myopic-RF has higher cost than
SC20-RF for all time periods except the second.
Finally, RL matches SC20-RF, within 1.2\%, for two periods and is the overall best realistic approach for four
of the six periods, with up to 17\% improvement over SC20-RF.

\begin{table*}[t]
	\centering
	\small
	\caption{Prediction results and classical machine learning metrics for the six approaches. The RL policy is the only approach that
		adapts to the current job characteristics, by more frequently performing mitigations when the cost of a UE
		would be high.}
	\begin{tabular}{@{}l@{}rrrrr@{~}rrr@{}}
		\toprule
		\textbf{Approach} & \textbf{TPs} & \textbf{FNs} & \textbf{FPs} & \textbf{TNs} & \multicolumn{2}{r}{\specialcellright{\textbf{Mitigations}\\\textbf{(TPs+FPs)}}} & \textbf{Recall} & \textbf{Precision}\\ 
		\midrule
		Never-mitigate & 0 & 67 & 0 & 259,228 & 0 & (0\%) & 0\% & \emph{n/a} \\
		Always-mitigate & 42 & 25 & 259,228 & 0 & 259,270 & (100\%) & 63\% & 0.02\% \\
		SC20-RF & 40 & 27 & 96,612 & 162,616 & 96,652 & (37\%) & 60\% & 0.04\% \\
		Myopic-RF 
		& 32 & 35 & 132,864 & 126,364 & 132,896 & (51\%) & 48\% & 0.02\% \\
		RL \\
		\quad MN4 job distribution & 26 & 41 & 43,544 & 215,684 & 43,570 & (17\%) & 39\% & 0.06\% \\ 
		\quad UE cost $<$ 100 node--hours & 18 & 49 & 49,678 & 209,550 & 49,696 & (19\%) & 27\% & 0.04\% \\
		\quad 100 $\leq$ UE cost $ < $ 1000 node--hours & 29 & 38 & 86,722 & 172,506 & 86,751 & (33\%) & 43\% & 0.03\% \\
		\quad UE cost $\geq$ 1000 node--hours & 41 & 26 & 240,155 & 19,073 & 240,196 & (93\%) & 61\% & 0.02\% \\
		Oracle & 42 & 25 & 0 & 259,228 & 42 & (0\%) & 63\% & 100\% \\ 
		\bottomrule
	\end{tabular}
	\label{tab:conf_matrix}
\end{table*}

\subsection{Different DRAM manufacturers}

\begin{figure}
	\centering
	\includegraphics[width=\columnwidth]{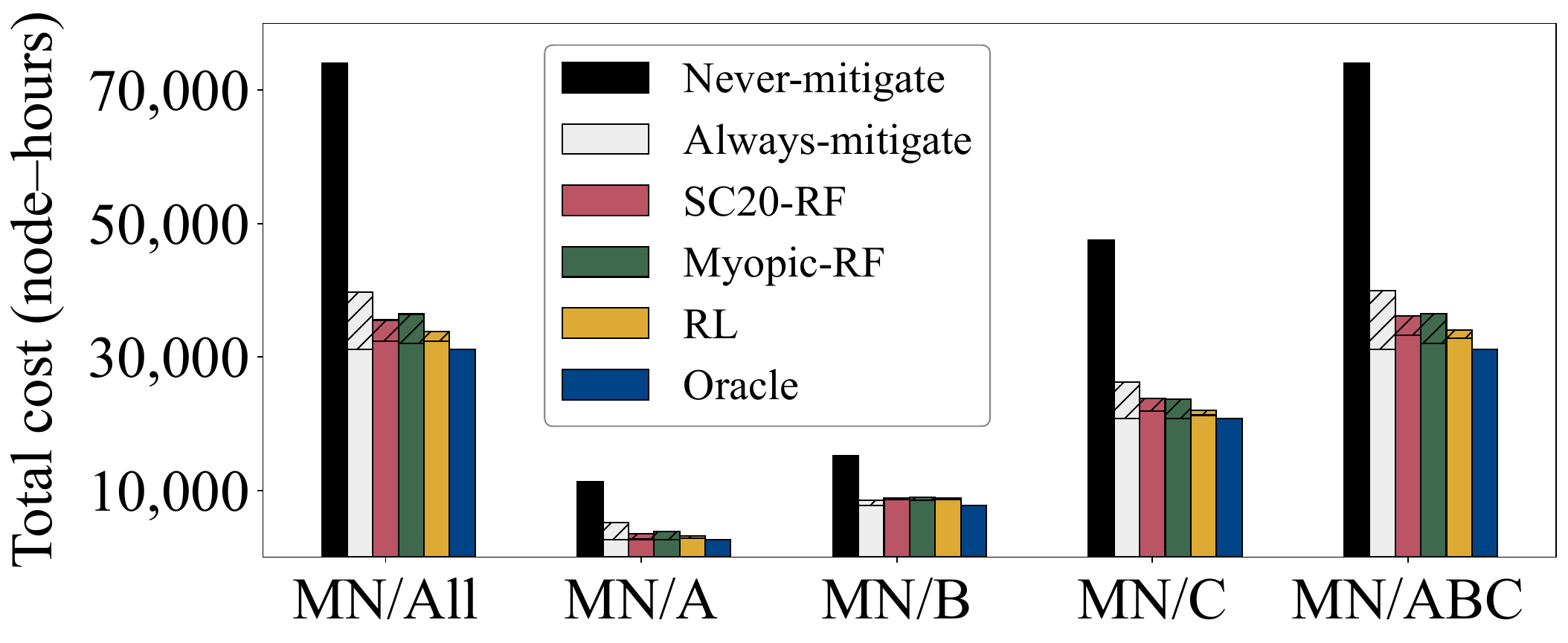}
	\caption{Total cost for the three anonymized DRAM manufacturers. Each bar is
		the sum of UE cost (solid color) and mitigation cost (with dashes). The RL agent has lower total
		cost than the other approaches due to a much lower mitigation
		cost. This result is consistent across all DRAM manufacturers.}
\label{fig:total-costs-manufacturers}
\end{figure}

\looseness -1 Figure~\ref{fig:total-costs-manufacturers} shows results for the three
anonymized DRAM manufacturers and a mitigation cost of two node--minutes. As before, the $y$-axis is the total cost across
the two-year production time period, as the sum of the costs of the UEs~(solid)
and mitigations~(dashed).  Separate results are shown for the whole system
(MN/All), as well as per manufacturer (MN/A, MN/B and MN/C) and their sum
(MN/ABC).

The relative effectiveness of the six approaches are broadly similar across all
scenarios considered: whether applied and evaluated to MareNostrum~3 as a whole
or separately to MN/A, MN/B and MN/C.  The MN/ABC results are similar to those
for MN/All, but slightly worse, likely because MN/All allows generalization
among manufacturers whereas MN/ABC does not.  We see that RL is significantly
better than SC20-RF for all scenarios except MN/B, for which the results are
similar.

\begin{figure}
	\centering
	\includegraphics[width=\columnwidth]{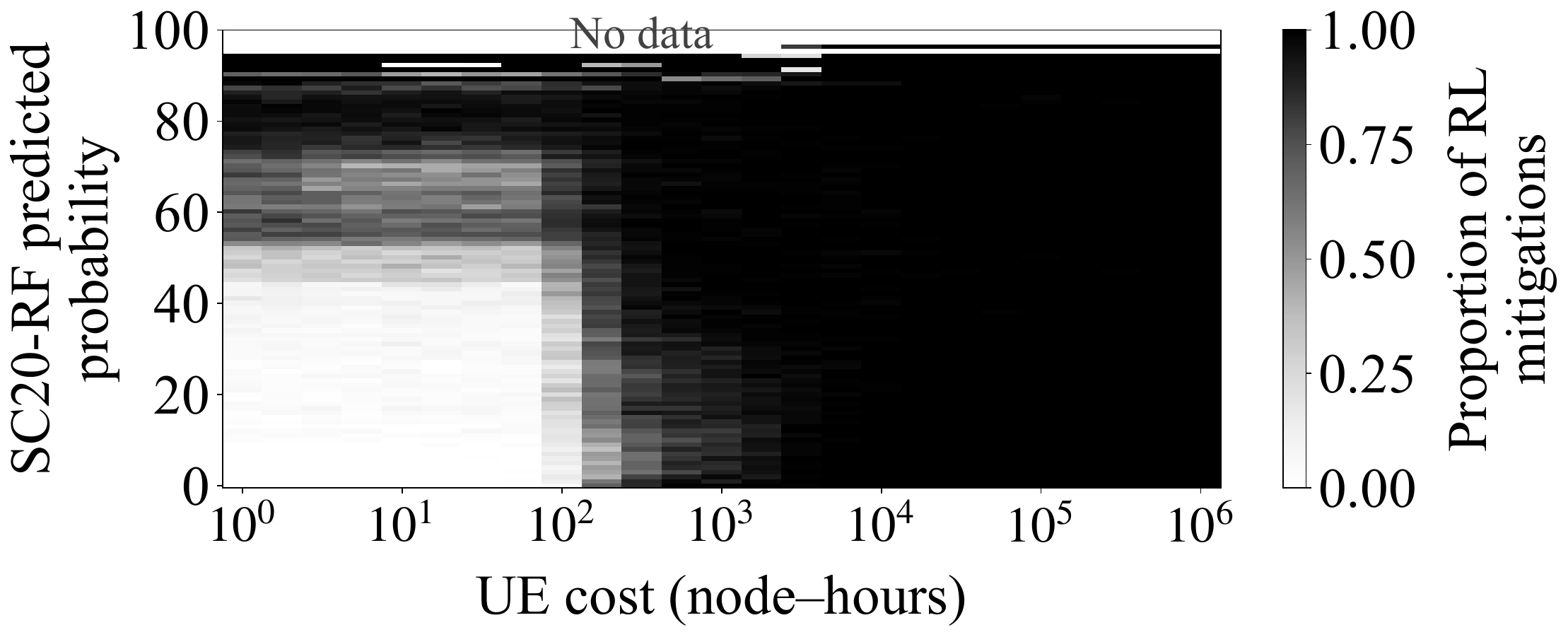}
	\caption{RL agent behavior: the shade indicates the fraction of events for which the
		agent triggers a mitigation as a function of the potential
		UE cost ($x$-axis) and the likelihood of a
		UE ($y$-axis) determined by the RF predictor.
	}
	\label{fig:RL-decisions}
\end{figure}

\subsection{RL agent behavior}
\label{ssec:rl-agent-behaviour}

Figure~\ref{fig:RL-decisions} illustrates the behavior of the RL agent, which
helps understand in which circumstances the agent decides to mitigate. The
$x$-axis (log scale) is the UE cost input to the RL agent (Equation~\ref{eqn:ue-cost}), which is the CPU time since the job start or the last checkpoint. The $y$-axis indicates the probability of a predictable UE.
Our agent has no such probability value, whether as an input, intermediate value or output.
As a proxy the $y$-axis therefore shows the probability output from the
state-of-the-art random forest predictor (SC20-RF). This is not an input to the
agent, but it does serve as a measure of the risk
of an upcoming UE. The shade in the chart indicates how often the agent
mitigates in those circumstances, with white meaning that it never
mitigates and black that it always mitigates. The region at the top marked ``No data'' indicates that
these datapoints are never seen during training. Each bin comprises
the relevant training episodes in the last cross validation split.

For low potential UE costs, below about 100 node--hours, and low predicted UE
probability, below about 50\%, the agent initiates few mitigations (bottom-left
part of the chart). As the UE cost increases, the proportion
of mitigations increases, with the agent usually triggering a mitigation when the UE cost is
above about 1000 node--hours, even if the likelihood of a UE is low.
Similarly, the agent usually initiates mitigations when the predicted probability is above
about 70\%. When the UE cost and likelihood of a UE are both high
(top-right part of the chart), the mitigation is almost always performed.

Figure~\ref{fig:RL-decisions} also shows that the agent properly generalizes to very large potential UE costs. The training data contains a small number of UE costs exceeding 1000 node--hours, with a maximum of 32,000 node--hours. Nevertheless, the agent consistently chooses to always mitigate for one and two orders of magnitude greater costs.

\subsection{Classical machine learning metrics}
\label{ssec:ues_predicted}

Table~\ref{tab:conf_matrix} shows the numbers of true positives~(TPs), false
negatives~(FNs),  false positives~(FPs) and true negatives~(TNs), as defined in
Section~\ref{ssec:classical-metrics}, for all six approaches. It also shows the
total number of mitigations, as well as recall and precision, which are derived
from these numbers.

\looseness -1 Never-mitigate never performs a mitigation action (positive), so it has zero
TPs or FPs.  Its recall, $TPs/(TPs+FNs)$, is therefore 0\% and its precision,
$TPs/(TPs+FPs)$, is undefined. Always-mitigate has 42 true positives,
corresponding to the 42 UEs that have at least one event in the 1-day
prediction window defined for the classical machine learning metrics.
But 25 UEs have no event in the 1-day prediction window, so
from the perspective of the classical machine learning metrics,
they cannot be mitigated and are counted as false negatives. This leads to a
recall of 63\%, which is the best possible for all approaches that perform a
mitigation only in response to an error event.  Always-mitigate,
however, has the lowest precision with a value of 0.02\%, due to the high
number of FPs (259,228).  SC20-RF has a slightly lower recall than
Always-mitigate, at 60\%, but many fewer FPs, leading to a significantly
improved precision of 0.04\%. Compared with SC20-RF, Myopic-RF has a
lower recall, at 48\% and a lower precision, at 0.02\%, due to the much lower
number of TPs.

To understand the behavior of the RL policy in different conditions, we show
separate results for the MareNostrum~4 job distribution (first row) and for
three uniformly randomly distributed ranges of UE costs (three remaining
rows). 
For the MareNostrum~4 job distribution (first row), the overall recall
is much lower than SC20-RF, at 39\%, while the precision is much higher, at
0.06\%. It may appear that the RL policy is making a poor tradeoff, due to the lower
number of mitigated UEs (true positives), but this is not the case, as
previously seen in the cost--benefit calculation. For UE costs less than 100
node--hours (second row), the agent requests mitigations only when there is either a high
probability of a UE or a high potential UE cost. This leads to the lowest recall of 27\% and a
precision of 0.04\%. While a higher recall would represent a greater number of
predicted UEs, the relatively small cost of any UEs would be insufficient to
justify the greater number of false positive mitigation actions.  For UE costs
between 100 and 1000 node--hours (third row), mitigation is performed more often,
leading to a recall of 43\% and a precision of 0.03\%. Finally, for UE costs
uniformly distributed between 1000 node--hours and the maximum job size of
32,000 node--hours (fourth row), RL behaves like Always-mitigate, leading to a
recall of 61\% (almost as high as 63\% for Always-mitigate) and a precision of
0.02\% (similar to Always-mitigate).
Finally, the Oracle has the highest possible recall, of 63\%, which is the same
as Always-mitigate, and the highest possible precision, of 100\%, since all
mitigation actions are true positives.

\looseness -1 With the exception of Oracle, which has the maximum value for both metrics,
recall and precision, it is not possible to conclude, from these metrics alone,
which policy is best. Firstly, precision considers TPs and FPs to have the same
weight, when in reality they differ in cost by orders of magnitude. Secondly,
there is generally a tradeoff between recall and precision, and an increase in
one typically results in a decrease in the other.  RL is the only policy that
dynamically adjusts this tradeoff to optimize the cost--benefit calculation.

\subsection{Job size sensitivity analysis}
\label{ssec:sensitivity}

\looseness -1

In order to verify the generality of our method on systems with different job
sizes, we repeat the experiment with up to ten times smaller or ten times
larger job sizes. Each experiment, for a different scaling factor, uses a
separately trained model, which corresponds to the normal use case of training
a model for the particular production system. The results are the average
across all six splits in the time series nested cross-validation, for the
complete system MN/All. In order to focus on the effect of job size, all
results maintain the same 2 node--minute mitigation time as before.

\begin{figure}
	\centering
	\begin{subfigure}[t]{\columnwidth}
		\centering
		\includegraphics[width=\columnwidth]{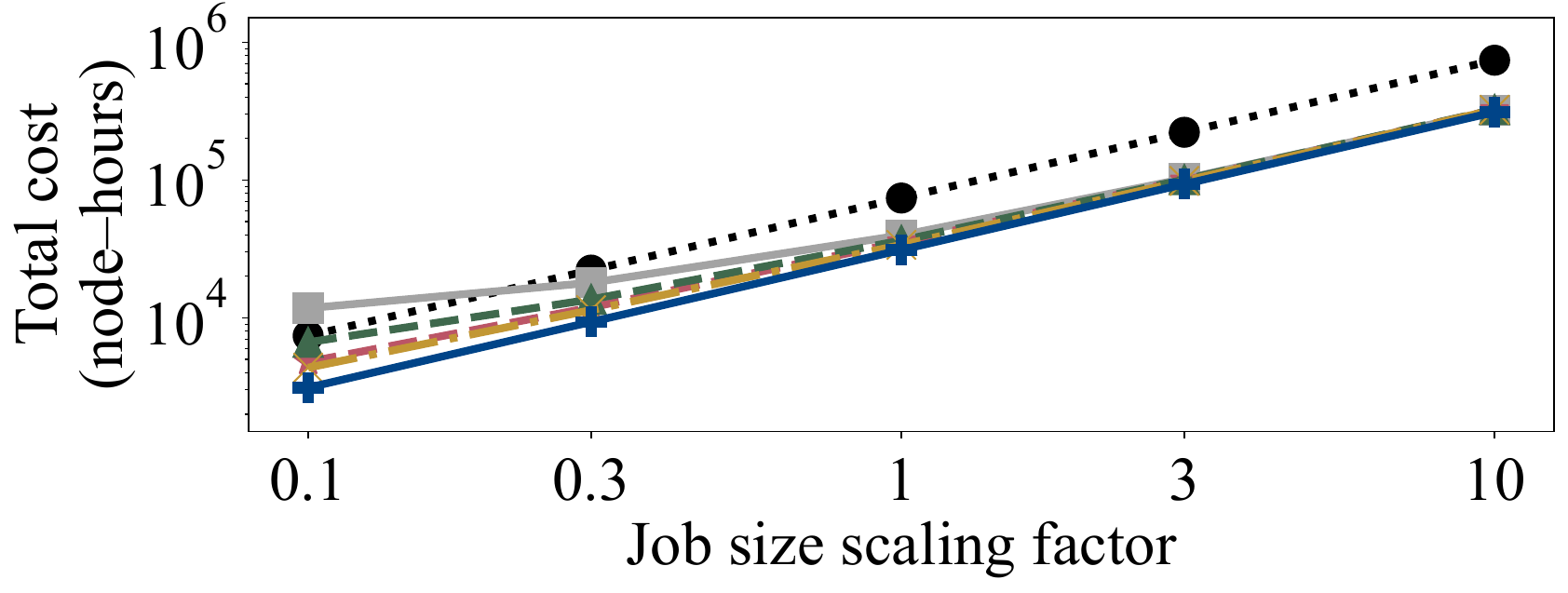}
		\caption{Total cost in node--hours (UE plus mitigation cost). The RL approach has the lowest total cost among all realistic approaches across the whole range. The best static policy changes from Never-mitigate to Always-mitigate at about one-third the job cost of MareNostrum.}
		
		\label{fig:total_cost_jobsX}
	\end{subfigure}
	\vspace{10pt}
	
	\begin{subfigure}[t]{\columnwidth}
		\centering
		\includegraphics[width=\columnwidth]{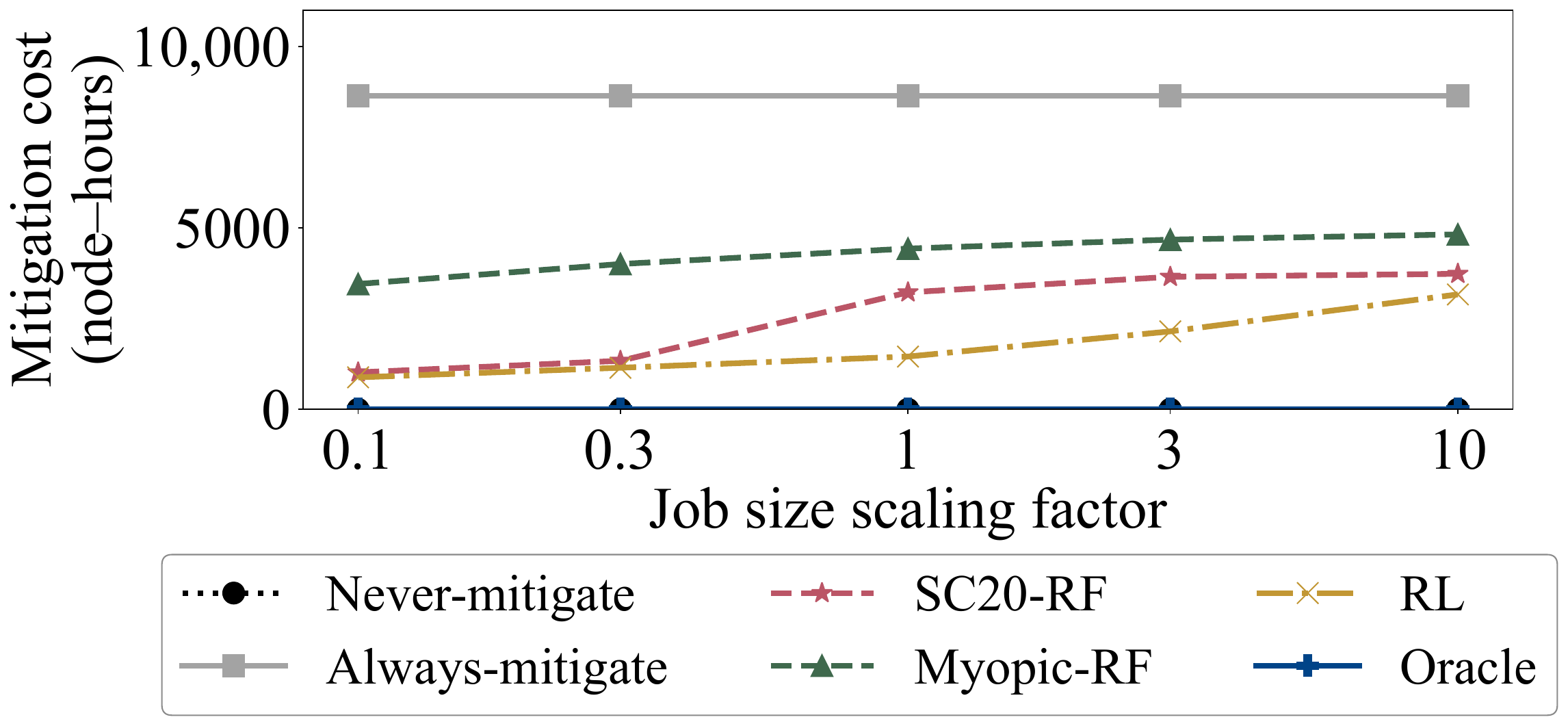}
		\caption{\looseness -1 Mitigation cost in node--hours. Never-mitigate, Always-mitigate and Oracle have fixed mitigation costs independent of the job sizes. SC20-RF adjusts to the job size through the external threshold parameter. Myopic-RF adapts to the job size scalefactor through the expected UE cost.  RL automatically adjusts to the job size.}
		\label{fig:mitigation_cost_jobsX}
	\end{subfigure}
	\caption{Sensitivity analysis for job size scaling factor.}
\end{figure}

\looseness -1 Figure~\ref{fig:total_cost_jobsX} shows the total cost of
mitigations and UE errors ($y$-axis, log scale), as a function of the job size
scaling factor ($x$-axis, also log scale).  A scaling factor of 1 corresponds
to the job distribution in the original MareNostrum 4 job log. As expected, the
cost of the uncorrected errors, and therefore the benefits of error mitigation,
both increase with the job scaling factor. Never-mitigate has total cost equal
to the UE cost, which is directly proportional to the scaling factor. For
scaling factors of 1, 3 and 10, Never-mitigate has total costs of 74,035,
222,104 and 740,346~node-hours respectively. Always-mitigate reduces the UE
cost by a factor of about 2.4, independent of the scaling factor, but it
adds a fixed mitigation cost of 8642 node--hours (see
Section~\ref{sec:results-cost-benefit}). For large scaling factors, its total
cost is dominated by the UE cost and is therefore close to proportional to the
scaling factor, about 2.4$\times$ lower than that of Never-mitigate. But
for job scaling factors less than about 0.2, the high mitigation cost of
Always-mitigate dominates and Never-mitigate becomes the best static baseline
policy.  On the large logarithmic scale covering two orders of magnitude,
SC20-RF, Myopic-RF and RL appear similar. All always perform better than the
static baseline policies, with the benefit of the prediction schemes largest
for moderately smaller job sizes than those of MareNostrum~4. The RL-triggered
mitigation policy has the lowest costs, of all policies except Oracle, at
32,391, 96,379 and 320,349 node-hours. 

Figure~\ref{fig:mitigation_cost_jobsX} focuses on just the mitigation costs
incurred by the different approaches, as the job sizes are scaled. The axes are
the same as for Figure~\ref{fig:total_cost_jobsX}, except that the $y$-axis now
shows only the mitigation cost, in node--hours, on a linear scale.  All three
prediction-based approaches, SC20-RF, Myopic-RF and RL adapt to the job size
scaling factor, by incurring a lower mitigation cost when the job sizes are
smaller. For SC20-RF, this adaptation is done by the user somehow specifying a
different optimal threshold parameter. In the case of Myopic-RF and the RL agent,
this adaptation is done automatically. Nevertheless,
we see that the RL agent consistently achieves a lower mitigation cost than the
other realistic approaches.

Various studies from NERSC, NSF and US national labs report typical HPC job sizes that
are two or three orders of magnitude larger than the jobs used in our
evaluation~\cite{hart2011deep, rodrigo2018towards, simakov2018workload,
moody2010detailed}.  Given that we have verified the generality of our method
with different job sizes, we argue that application of our method in these
systems would lead to roughly proportional savings that are two to three orders
of magnitude higher than our results for MareNostrum~3.

\looseness -1 Numerous studies considering large-scale HPC systems report checkpointing
overheads of tens of percents of the overall available node--hours~\cite{wang2010hybrid, daly2008application, elliott2012combining, oldfield2007modeling, bautista2011fti, schroeder2007understanding, elnozahy2009system, bergman2008exascale, sato2012design},
which is considerably higher than observed in our study.
Driven by the needs of large-scale systems,
optimizing the checkpointing interval to reduce overheads is an active area of
research.  To the best of our knowledge, our study is the first to propose and
evaluate an RL adaptive mitigation scheme that would lead to significant system
performance gains in HPC systems with a high failure mitigation cost.

\subsection{Application completion times}
\label{sec:ExeTimes}

\looseness -1 Our study focuses on the system-level impact of DRAM failures and the
potential for the system operator to reduce the number of lost node--hours. At the same time, error mitigation approaches can also improve the
service delivered to individual users. Wang et al.~\cite{wang2010hybrid} and
Elliott et al.~\cite{elliott2012combining} report an $n$-fold reduction in 
job wall-clock time,
due to the deployment of checkpointing. The trade-off
between false positives and false negatives
 motivates the exploration of adaptive mitigation policies that
adjust to the job and system state.  An interesting avenue of future work would
be to analyze the application-level benefits of our proposal, considering
application completion time and its variability under different mitigation
approaches.

	\section{Related work}
	\label{sec:Related-work}

\subsection{Corrected DRAM errors}

\looseness -1 Most of the earlier studies on the classification and prediction of memory
errors focus on corrected DRAM errors~\cite{Costa:ProactiveMemory-ErrorAvoidance, Baseman:DRAMFaultCharacterization, Baseman:DSN17, Sun:FailurePredictionUsingDL, Du:FailurePredictionUsingOnlineLearning, Nie:HPCA2016, Nie:DSN2018}.
These studies help to understand corrected DRAM error rates and 
distributions, and they identify correlations with various features
that are useful for prediction. It is important, however, to determine how and
to what extent this information could contribute to measurable improvements
in system reliability.  System reliability is only impacted by UEs~\cite{Schroeder:TDSC2010, Martino:DSN2014, Gupta:SC2017,
Zivanovic:MEMSYS2019}, and there is no direct relation between corrected and
uncorrected error prediction~\cite{Schroeder:SIGMETRICS2009, Sridharan:SC2012,
Giurgiu:Middleware2017, Levy:SC2018, Zivanovic:MEMSYS2019,
radojkovic2020towards}.  It is therefore challenging to modify an existing
CE predictor to predict UEs~\cite{Du:FailurePredictionUsingOnlineLearning}.

\subsection{Corrected vs.\ uncorrected DRAM errors}

A few studies show that the probability of an uncorrected DRAM error is higher if the DIMM previously experienced corrected errors~\cite{Schroeder:SIGMETRICS2009, Sridharan:SC2012, Zivanovic:MEMSYS2019}. 
This reasoning  is used by system protection mechanisms that try to prevent future uncorrected
errors using simple heuristics
to retire potentially failing memory pages \cite{Hwang:ASPLOS2012, Meza:DSN2015, Tang:DSN2006, Lenovo:PlanningGuide:2018, Du:ICCD2021} or replace the affected DIMMs \cite{HP:ProLiantDL580-UserGuide:2016, Intel:SEL-TroubleshootingGuide:2017, Martino:DSN2014, Schroeder:SIGMETRICS2009, Du:EDCC2020}. 
The recent large-scale study of Cheng et al.~\cite{Cheng:SRDS2022}, however concludes that UE failures are hard to predict, since typically only a small number of CEs occur before these failures and the CEs only manifest within a short time before the failures happen.

\subsection{Uncorrected DRAM errors}
The community has more recently applied advanced machine learning methods to predict uncorrected DRAM errors.

\looseness -1 In a 2017 paper, Giurgiu et al.~\cite{Giurgiu:Middleware2017} present the first machine learning model to predict uncorrected DRAM errors. Their random forest model is based on preceding corrected errors and
measurements from over 100~sensors that monitor the system. 
The model is designed and evaluated based on event logs from 49,800 IBM servers in multiple geographical locations during a period of over three years. 

\looseness -1 Mukhanov et al.~\cite{mukhanov2019workload} explore the importance of workload characteristics, such as IPC and memory bandwidth utilization, on the characterization and prediction of
DRAM errors. 
The study is performed on a single server with 72~DRAM chips running under non-nominal circuit parameters: scaled refresh period, lowered voltage and increased temperature.

\looseness -1 Workload-aware DRAM error prediction is further explored by Wang et al.~\cite{wang2021workload} in a large-scale study
of a cloud datacenter comprising 382,608 servers. 
three tree-based models.
The results show that considering the workload's used memory bandwidth and latency improves upon a baseline prediction based only on the platform characteristics, the number of CEs and their location.  

\looseness -1 These error prediction models are evaluated using standard evaluation metrics, such as precision and recall.  
Such metrics allow comparison with the state of the art in machine learning,
but they do not allow us to conclude how well the predictor would serve to
reduce the costs of UE errors.  
Metrics like precision put the same weight on multiple different prediction outcomes (true/false positives/negatives), whose costs differ by orders of
magnitude, and the recall metric ignores false positives, which incur costly mitigation measures. Overall,
these standard data prediction metrics are insufficient to evaluate HPC failure predictors, and their use should be complemented with a cost–benefit analysis. 

\looseness -1 Boixaderas et al.~\cite{boixaderas2020cost} is the first study that performs
such an analysis. It compares the system resources needed for model training, failure
prediction and mitigation against the compute time that is saved by
successful failure mitigation. The study develops and compares six machine
learning classifiers and it proposes an error prediction method based on random forest. 
The method is trained and evaluated using logs from the MareNostrum supercomputer.

\looseness -1 Two recent studies from Internet servers and the cloud domain use a similar cost--benefit calculation
for DRAM failure prediction.
Li et al.~\cite{li2022correctable} analyze DRAM errors from a ByteDance Internet facility comprising around 100,000 Intel SkyLake and Cascade Lake servers. 
The study proposes three simple UE predictors based on the CE history and the
details of the system's ECC algorithms 
to predict the risky 
CEs.
The ECC can correct risky CE patterns with limited assured coverage, so a small variation in a risky CE will likely result in a UE.
The authors show cost--benefit improvements when risky CE patterns are considered.

\looseness -1
Zhang et al.~\cite{zhang2022predicting} predict the loss of node availability due to DRAM failures for the Alibaba Cloud Elastic Compute Service with more than half a million nodes. 
In addition to UEs, the study also analyzes node unavailability caused by CE storms, high CE rates
that saturate system error handling mechanism and cause a node to become unresponsive.  
The authors predict DRAM-caused system failures based on a 18 heuristic rules combined with four binary classification
ML models.

The three previous studies create static prediction models that don't adapt to the current state of the system.
This is problematic in high-performance computing because the failure cost 
varies among jobs whose size and duration can differ by orders of magnitude. 
Our method dynamically
adapts to the current system state, learning a policy that takes account
of the running job characteristics.

	\section{Conclusions}
	\label{sec:Conclusions}
	
\looseness -1 This paper presented and evaluated a reinforcement learning
method that anticipates and triggers the mitigation of DRAM uncorrected errors.
The method is trained and evaluated using two years of production error and job
logs from the MareNostrum supercomputer. Our cost--benefit analysis shows that
the adaptive mitigation method reduces the lost compute time by 54\%, an
overall saving of 20,000 node--hours per year. By adapting to the current job
size, rather than using a static cost estimate, the saved node--hours is just
6\% below the optimal Oracle method, reducing the distance from optimal by more
than a third compared with the state-of-the-art random forest method. We verify
the generality with different job sizes and argue that application of our
method on larger HPC systems with larger HPC job sizes would lead to roughly
proportional savings two to three orders of magnitude higher than observed on
MareNostrum.

All source code is released as open source. The only user-defined parameters are the total mitigation cost and whether the job can be restarted from a mitigation point. The method
can therefore be applied to other systems without customization or tuning.
We would encourage the community to
evaluate and further improve the method on their systems and share their
findings.



	\begin{acks}
		The work was supported by the Spanish Government, under the contracts PID2019-107255GB-C21 and CEX2021-001148-S funded by MCIN/AEI/ 10.13039/501100011033. The work also received funding from the Department of Research and Universities of the Government of Catalonia to the AccMem Research Group (Code: 2021 SGR 00807). Paul Carpenter holds the Ramon y Cajal fellowship RYC2018-025628-I funded by MICIU/AEI/10.13039/501100011033 and "ESF Investing in your future".
	\end{acks}
\bibliographystyle{ACM-Reference-Format}
\bibliography{ue_rl}

%
%
%
%
%
%
%
%

\end{document}